# Bose-Einstein condensate as a diagnostic tool for an optical lattice formed by 1064 nm laser light

V.V. Tsyganok [1], D.A. Pershin [1,2], V.A. Khlebnikov [1], D.A. Kumpilov[1,3], I.A. Pyrkh[1,4], A.E. Rudnev[1,3], E.A. Fedotova[1,3], D.V. Gaifudinov[1,3], I.S. Cojocaru[1,2], K.A. Khoruzhii[1,3], P.A. Aksentsev[1,4], A.K. Zykova[1] and A.V. Akimov[1,2,5]

[1]*Russian Quantum Center, Bolshoy Boulevard 30, building 1, Skolkovo, 143025, Russia*

[2]*PN Lebedev Institute RAS, Leninsky Prospekt 53, Moscow, 119991, Russia*

[3]*Moscow Institute of Physics and Technology, Institutskii pereulok 9, Dolgoprudny, Moscow Region 141701, Russia*

[4]*Bauman Moscow State Technical University, 2nd Baumanskaya, 5, Moscow, 105005, Russia*

[5]*National University of Science and Technology MISIS, Leninsky Prospekt 4, Moscow, 119049, Russia*

email: a.akimov@rqc.ru

Recently, the thulium atom has been cooled down to the temperature of Bose-Einstein condensation. While the condensate of the thulium atom has a lot of applications in quantum simulations and other areas of physics, it can also serve as a unique diagnostic tool for many atomic experiments. In the present study, the Bose-Einstein condensate of the thulium atom was successfully utilized to diagnose an optical lattice and detect unwanted reflections in the experiments with the 1064 nm optical lattice, which will further be used in a quantum gas microscope experiment.

## I. INTRODUCTION

Ultra-cold atoms find more and more applications in sensing [1, 2], time keeping [3–8], quantum computations [9–11], and quantum simulations [12–15]. The latter typically start with achieving a quantum degeneracy state. Among novel atoms that have just recently joined the family of atoms with achieved degeneracy, there is the thulium atom [16], which possesses a lot of non-chaotic Fano-Feshbach resonances [17, 18] at a low field and has a magnetic moment of $4\mu_B$ in the ground state.

One of the effective approaches to quantum simulation is using a quantum gas microscope [15, 19–24], which requires the formation of an optical lattice (OL). While working on the construction of such a device, there is a difficulty in the diagnostics of exact beam geometry. While this question is solvable in principle, it is always useful to have an in-chamber probe, which can measure the actual configuration in a vacuum without opening the vacuum chamber. The Bose-Einstein condensate (BEC) is indeed a useful sensor for the configuration of laser beams. The configuration of beams in the chamber could be imprinted on the matter wave of condensate in position space via the Kapitsa-Dirac (or Raman-Nath) effect [25–28]. The accumulated phase modulation could be transferred into momentum if condensate is released. In the case of a periodic potential

$$U(z) = U_0 \sin^2(k_L z), \tag{1}$$

where $U_0$ is the depth of the OL, $k_L = |\vec{k}_1 - \vec{k}_2|/2 = \pi/\lambda_L$ is one-half of the magnitude of the reciprocal lattice vector, $\vec{k}_1$ and $\vec{k}_2$ are wave-vectors of OL beams, $\lambda_L = \lambda/(2\sin(\alpha))$ is a lattice period, $2\alpha = (\vec{k}_1; \vec{k}_2)$.

As atoms are exposed to the periodic potential, they gain additional momentum, which could be detected during BEC expansion. The momentum gained by the atoms in BEC is

$$\vec{p}_m = 2m\hbar \vec{k}_L, \ m = 0, 1, 2..., \tag{2}$$

where $\hbar$ is a Planck constant, and a fraction population of the $m$-th diffraction order. The population of the $m$-th diffracted order with $\vec{p}_m$ in the Raman-Nath regime [28] is a function of the exposure on the potential modulation $T_{pulse}$ and OL depth $U_0$ as:

$$P_m = J_m^2(U_0 T_{pulse}/2\hbar), \tag{3}$$

where $J_m$ are Bessel functions of the $m$-th kind. Thus, the momentum gained depends on both the exposure time and periodic potential depth. If the exposure time is known, the magnitude of the gained momentum carries information on the depth of the periodic potential. Moreover, the variation of the exposure time makes it possible to vary the sensitivity of the BEC to the potential depth, thus making it possible to focus on the specific range of depth.

The present study successfully utilized the Kapitsa-Dirac effect on thulium atoms BEC to diagnose the OL and detect unwanted reflections in the experiments with the 1064 nm OL. The measurements were then verified by measurements of trap frequencies, calculated based on the results of the Kapitsa-Dirac experiment.

## II. KAPITSA-DIRAC EXPERIMENT

The details of the experimental setup could be found elsewhere [16, 29–33]. An atomic cloud was precooled in a magneto-optical trap, operating at $4f^{13}\left(^{2}F^{o}\right)6s^{2} \rightarrow 4f^{12}\left(^{3}H_{6}\right)5d_{5/2}6s^{2}$ (530.7 nm) transition with precooling at $4f^{13}\left(^{2}F^{0}\right)6s^{2} \rightarrow 4f^{12}\left(^{3}H_{5}\right)5d_{3/2}6s^{2}$ (410.6 nm) transition. Then atoms were loaded in a far-detuned magneto-optical trap [34] to form a cloud polarized to the lowest magnetic sublevel of the ground state $|F=4; m_{F}=-4\rangle$. The cloud temperature after polarization was measured by the time-of-flight technique to be about 13 µK. Then atoms were loaded into a 532.07 nm optical dipole trap (ODT) with a waist size $w_{z}=15.8$ µm and $w_{y}=25.7$ µm as described in [33] and forced evaporatively cooled by decreasing the power of 532-laser beams [16]. Once BEC was achieved following the sequence described in [16], the atoms were used for further experiments.

The OL was formed by 1064 nm beams (see Figure 1), focused into the region of the ODT with beam waists of $w_{0}=88\pm4$ µm (at the e$^{-2}$ level by power). Beams were crossed at an angle of $2\alpha=32°$. The configuration is dictated by the geometry of the vacuum chamber (Kimball Physics MCF800-ExtOct-G2C8A16).

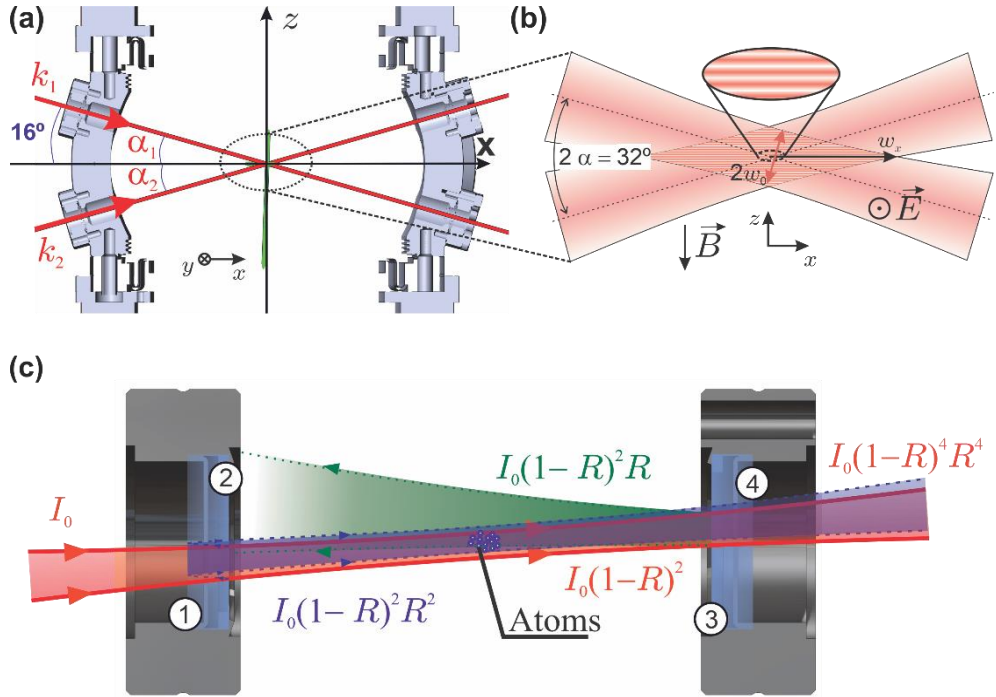

*Figure 1. (a) – 1064 nm beams forming an optical lattice. (b) – Details of the 1064 nm optical lattice. The polarization of the light is out of the plane of the figure. (c) – Reflections of one beam of the optical lattice tilted at a slight angle to the surface 1. 1, 2, 3, 4 – the number of surfaces of the viewport. Red solid lines correspond to input and output beams with no reflection. Violet dashed lines correspond to the forward propagating reflection from the input viewport. Green dotted lines correspond to a reflected beam from the second viewport.*

BEC served two purposes: first, it was used to check the presence of a 1064 nm OL, and second, to diagnose unwanted reflections, which unfortunately were present due to the absence of an antireflection coating. Viewports that were used in the OL setup had an anti-reflection coating for 532 nm but not for 1064 nm. The reflection coefficient $R$ of the window was verified using a spare window and was found to be $R = 15.5 \pm 1.7\%$ using the reflection of the same laser beam used for the OL. This reflection coefficient is significant and thus unwanted reflections may significantly modify the parameters of an OL.

To form the OL, the beams were adjusted in two steps: first, both beams were aligned to be perpendicular to the vacuum chamber viewports (see Figure 1a) by matching the reflection from both surfaces of the port and the beam. Then the beams were intersected inside the vacuum chamber and moved to the location of BEC (formed at 532 nm ODT) by observing the reloading of the atomic cloud into a 1064 nm OL. Once beams were aligned perpendicularly to the viewport, the angle of the beam could not be changed significantly. The geometry of the setup which has fused silica viewports with a view area of 16 mm and a thickness of 1.8 mm with a distance between those of 221.3 mm only makes it possible to change the angle of the beam by no more than ±3.7° degrees (taking into account the waist of the beam on the window at the level e$^{-5}$ of 0.7 mm). The viewports themselves are located 16° with respect to the horizontal $xy$-plane.

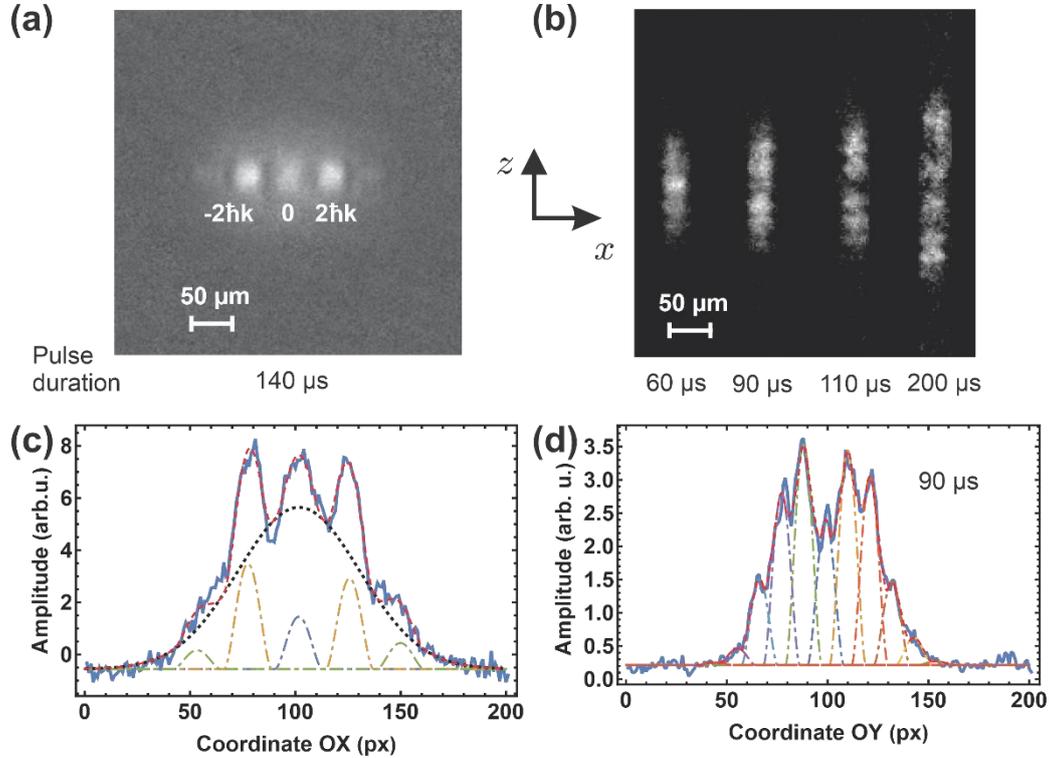

*Figure 2. a) 15 ms expansion of BEC after 140 μs-pulse of a single OL beam with a power of 0.7 W. b) 15 ms expansion of BEC after 60, 90, 110, and 200 μs-pulse of the OL. c) and d) Atomic distribution with the fit for figures b and c (90 μs). The red dashed line is a fit with eq.(4) with i=2 for (c) and i=5 for (d). The dotted black line is a fit of the thermal cloud. Dotted-dash lines are the fit of diffraction peaks.*

Due to the alignment procedure, there is a good chance of strong overlap of the OL beam and its forward reflection from the viewport for each of the beams (beam limited by violet arrows Figure 1c), forming the OL. Besides, there is a possibility of the reflection from the second viewport (green beam, Figure 1c).

The "green" reflection will expand between surfaces 3 and 2 considerably but in the position of the OL region, the expected beam size is still about 830 μm and is thus not negligible. Moreover, with this width, the reflected beam would overlap with the OL region for any angle $15.7° < \alpha < 16.3°$. At the same time, the "violet" reflection (reflection from the first surface) would overlap with the OL region for the angle $13.1° < \alpha < 18.9°$. All these reflections may significantly modify the parameters of an OL.

Exact geometrical measurements, while possible, may be quite challenging for the vacuum volume which is part of the optical setup. Therefore, the beam geometry may be analyzed using the interference of BEC using the Kapitsa-Dirac effect. This analysis was performed first for a single beam (Figure 2a) and then for an OL (Figure 2b).

To check the self-reflections of individual beams, thulium atoms were first cooled to the BEC state at the 532 nm ODT. Then, while ODT was still on, one OL ($\vec{k}_1$ in Figure 1a) beam of interest was turned on for 140 μs. The power of one ODT beam was 1.1 W before the vacuum chamber. Once the OL beam was turned off, the 532 nm ODT was turned off as well, and the cloud was expanding freely for 15 ms. After expansion interference, fringes, corresponding to the scattering of BEC on the periodic potential, were observed (see Figure 2a). The $m$-spot picture corresponds to the momentum, gained by BEC while interacting with the OL beam, namely $\left|2m\hbar\vec{k}_l\right\rangle$, $m = 0, 1, 2...$, where $\vec{k}_l$ is the wave vector of the periodic lattice, present in the beam due to the self-reflection of the beam. Figure 2c shows the fit of the one-dimension atomic distribution with ($i = 2$):

$$N(x) = bg + \frac{N_t}{\sigma_x \sqrt{\pi}} Exp\left[-\frac{(x-x_0)^2}{\sigma_x^2}\right] +$$
$$+ \begin{cases} \sum_{m=-i}^{i} \frac{15 N_m}{16 R}\left(1 - \frac{(x-x_0+m\Delta)^2}{R^2}\right)^2, & x-x_0+m\Delta \leq R \\ 0, & x-x_0+m\Delta \geq R \end{cases} \qquad (4)$$

where $bg$ is a background, $N_t$ is the number of atoms of the thermal part, $\sigma_x$ is a cloud size of the thermal cloud, $x_0$ is the position of the center of mass, $N_m$ is the number of atoms of the $m$-th diffraction order, $R$ is a BEC radius, $\Delta$ is a distance between the nearest diffraction orders. The value of $\Delta$ could also be calculated as:

$$\Delta = \frac{2\hbar k_L}{M_{Th}} t, \qquad (5)$$

where $M_{Th}$ is the thulium mass, $t$ is the time of flight. According to the fit, $\Delta_{experiment} = 47 \pm 2$ μm and is in good agreement with theory (eq. (5)) where $\lambda_L = 532$ nm

$\Delta = 46$ μm. Therefore, the angle for that beam is $15.7° < \alpha < 16.3°$, and it is necessary to take into account all two reflections.

The second beam with $\vec{k}_2$ (Figure 1a) does not make diffraction pictures of BEC at any power. Therefore, one can exclude that the geometry of this beam makes an intersection with self-reflection from surfaces 3 and 4 (Figure 1b) and the angle for that beam could be $13.1° < \alpha < 15.7°$ and $16.3° < \alpha < 18.9°$.

Also, both 1064 nm beams on the OL lattice should be with quite different wave vectors $\vec{k}_l^*$ which should be oriented vertically (along the $z$ axis) in the lab frame (see Figure 1b). Indeed, if the Kapitsa-Dirac type of experiment is repeated with both beams forming the OL, one could see BEC diffraction peaks in a vertical direction (Figure 2c). In this case, the OL was turned on only at 60, 90, 110, and 200 μs since the interference patent was expected to have a large depth (the power at each OL beam was 0.54 W before the vacuum chamber), and therefore less exposure was needed. Reduced exposure also significantly reduced the visibility of the horizontal peaks. Figure 2d shows a one-dimension fit with eq.(4) (i=5) of the data corresponding to the 90 μs exposition of the OL. The value of $\Delta_e = 17 \pm 2$ μm obtained from the fit is also in good agreement with theory $\Delta = 17$ μm for $\lambda_L = 1064 \text{ nm} / 2\sin[16°] = 1930$ nm (see eq.(5)).

## III. MODELING

The result of the measurements showed that only one of the beams forming OL had reflection from the second viewport (surface 3), marked as green in Figure 1c. This is perfectly within what is possible in the setup, but hard to check directly. To model the potential, only two reflections for one beam $\vec{k}_1$ (see Figure 1c) and one reflection for beam $\vec{k}_2$ were used, as the next order of reflection does not make a significant change. The full value of the electric field $\vec{E}_{SUM}$ in the cross region in this model would be:

$$\vec{E}_{SUM} = \vec{E}_{10} + \vec{E}_{11} + \vec{E}_{13} + \vec{E}_{20} + \vec{E}_{21}, \quad (6)$$

where the first index is the number of the OL beams, and the second index is the number of reflecting surfaces as those marked in Figure 1c. Here 0 stands for no reflection. The relative amplitudes of the transmitted and reflected beams may be calculated as it is indicated below. As the reference frame, the one indicated in Figure 1b was used.

$$\vec{E}_{10} \propto \vec{E}_{20} \propto \left\{\sqrt{\frac{2I_0(1-R)^2}{\varepsilon_0 c}}, 0, 0\right\} \frac{1}{\sqrt{1+\left(\frac{\lambda x_{1,2}}{\pi \omega_0^2}\right)^2}} \times$$

$$\times Exp\left[-\frac{(z_{1,2})^2 + y^2}{\omega_0^2 \left(1+\left(\frac{\lambda(x_{1,2})}{\pi \omega_0^2}\right)^2\right)}\right] Exp\left[i(\vec{k}_{1,2}, \vec{r})\right], \quad (7)$$

$$\vec{E}_{11} \propto \vec{E}_{21} \propto \left\{\sqrt{\frac{2I_0(1-R)^2 R^2}{\varepsilon_0 c}}, 0, 0\right\} \frac{1}{\sqrt{1+\left(\frac{\lambda(x_{1,2}+\delta_1)}{\pi \omega_0^2}\right)^2}} \times$$

$$\times Exp\left[-\frac{z_{1,2}^2 + y^2}{\omega_0^2 \left(1+\left(\frac{\lambda(x_{1,2}+\delta_1)}{\pi \omega_0^2}\right)^2\right)}\right] Exp\left[i\left(\vec{k}_{1,2}, \vec{r} + \frac{2\pi}{\lambda}\delta_1\right)\right], \quad (8)$$

$$\vec{E}_{12} \propto \left\{\sqrt{\frac{2I_0(1-R)^2 R}{\varepsilon_0 c}}, 0, 0\right\} \frac{1}{\sqrt{1+\left(\frac{\lambda(x_{12}+\delta_2)}{\pi \omega_0^2}\right)^2}} \times$$

$$\times Exp\left[-\frac{z_{12}^2 + y^2}{\omega_0^2 \left(1+\left(\frac{\lambda(x_{12}+\delta_2)}{\pi \omega_0^2}\right)^2\right)}\right] Exp\left[i\left(\vec{k}_{12}, \vec{r} + \frac{2\pi}{\lambda}\delta_2\right)\right], \quad (9)$$

where $x_1 = x\cos[\alpha] - z\sin[\alpha]$, $x_2 = x\cos[\alpha] + z\sin[\alpha]$, $z_1 = x\sin[\alpha] + z\cos[\alpha]$, $z_1 = -x\sin[\alpha] + z\cos[\alpha]$, $x_{12} = -x_1$, $z_{12} = -z_1$, $\vec{k}_{1,2}, \vec{k}_{12}$ are wave-vectors of beams and reflected beams, $\delta_1 = 2nd\Big/\sqrt{1-(\sin[\alpha-16°]/n)^2}$ is the optical path difference arising in the window with the index of refraction $n$ and length $d$ and $\delta_2 = D$ is the optical path difference arising in the vacuum chamber with a diameter $D$.

The depth of the OL is:

$$U_{OL}(\vec{r}) = -\frac{2\pi a_B^3}{c}\operatorname{Re}(\alpha_{tot})I(\vec{r}),$$
$$I(\vec{r}) = \frac{\varepsilon_0 c}{2}\left|\vec{E}_{SUM}(\vec{r})\right|^2,$$
(10)

where $\varepsilon_0$ is vacuum permittivity. In the simulations, the input fields of both beams forming the OL were assumed to be the same. The simulated beam OL profile is indicated in Figure 3a. Here, the value for polarizability was taken from [35], and the field magnitude was found from the experimentally measured beam powers and experimentally found beam waste $\omega_0 = 88 \pm 4$ μm.

To test the modeled OL structure, one could perform measurements of the trap frequency $(\nu_x, \nu_y, \nu_z)$ (see below). In a parabolic approximation, to find trap frequencies, one needs to calculate the second derivative of the potential:

$$\nu_i = \frac{1}{2\pi}\sqrt{\frac{1}{m_{th}}\frac{\partial^2 U_{OL}}{\partial i^2}},\ i \in \{x, y, z\}.$$
(11)

## IV. TRAP FREQUENCIES IN THE OPTICAL LATTICE

To measure the trap frequencies of the OL, formed by a laser beam with a wavelength of 1064 nm, the following experiment was performed. First, atoms were loaded into a 532 nm ODT without waiting for the end of evaporation cooling. Then the power of 1064 nm OL laser beams was increased up to 4.0 W, 5.3 W, or 5.9 W per beam in a vacuum chamber (depending on the experiment) to reload atoms into the OL (see Figure 3b). The time duration of this stage was 2.2 s. Typically, about $1.0 \times 10^6$ atoms with a temperature of 10 μK were loaded into the OL. The magnetic field during the entire stage remained constant and was equal to $-3.91$ G along the Oz-axis (Figure 1a,b). An atomic cloud was then held in the OL for about 1 s, and then the measurement step was performed.

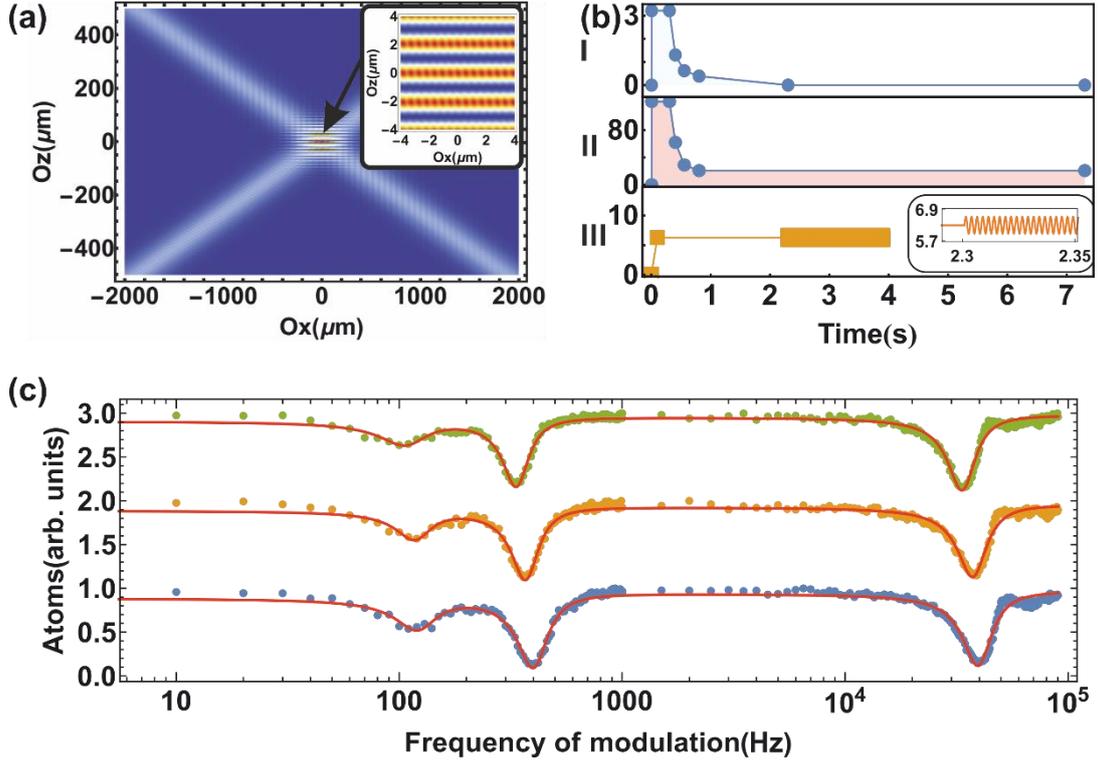

*Figure 3. a) Reconstructed potential at the Oxz-plane, b) Pulse scheme of the parametric heating experiment. I – power of the horizontal beam ODT beam in Watts, II – ODT beam scanning in μm, III – power per one OL beam. c) Atom loss during 1.6 s of power modulation of 1064 nm laser beams. Green, orange, and blue dots correspond to experimental data with power at one beam of 4.0 W, 5.3 W, and 5.9 W. Lines fit with three Lorentzian functions of experimental data.*

To measure the frequency of the OL, the parametric heating of an atomic cloud was used. Such heating was implemented by the modulation of both 1064 nm laser beams (see inset in Figure 3b). The OL modulation was turned on for about 1.6 s using an acousto-optic modulator with a small amplitude of 5% power of beams. Once modulation was completed, the remaining number of atoms was detected. The experiment was repeated for various modulation frequencies, resulting in a figure, demonstrated in Figure 3c.

The trap frequencies $\nu_i$ were determined from the minimum of the obtained curve $\nu_{OL}$ as:

$$\nu_i = \frac{\nu_{OL}}{2} \qquad (12)$$

were $i$ represents $x, y, z$. The measured frequencies were found to be (see Figure 3c):

Table 1. OL frequencies from experimental data.

| Power | 4.0 W | 5.3 W | 5.9 W |
|---|---|---|---|
| $\nu_x$ | 52±3 Hz | 58±3 Hz | 61±3 Hz |
| $\nu_y$ | 167±2 Hz | 184±2 Hz | 199±2 Hz |
| $\nu_z$ | 16,699±94 Hz | 18,616±121 Hz | 19,640±199 Hz |

As $\nu_z$ has the most sensitivity to the tensor part of the polarizability, the shift of parametric resonance is measured to check tensor polarizability. The measurements were performed for two different orientations of the magnetic field. The first one was done for the magnetic field $-3.91$ G oriented along the Oz-axis, and the second one for the magnetic field 3.91 G oriented along the Oy-axis (Figure 4a). The resulting frequencies were found to be 40,018±89 Hz and 39,484±77 Hz, correspondingly. Thus, the polarizability ratio $\alpha_{VERT\,B}/\alpha_{HOR\,B} = \left(\nu_{Z\,VERT\,B}/\nu_{Z\,HOR\,B}\right)^2$ was found to be about 0.97±0.03 from the experiment, which is close to previous experimental results 0.97±0.3 [7, 35].

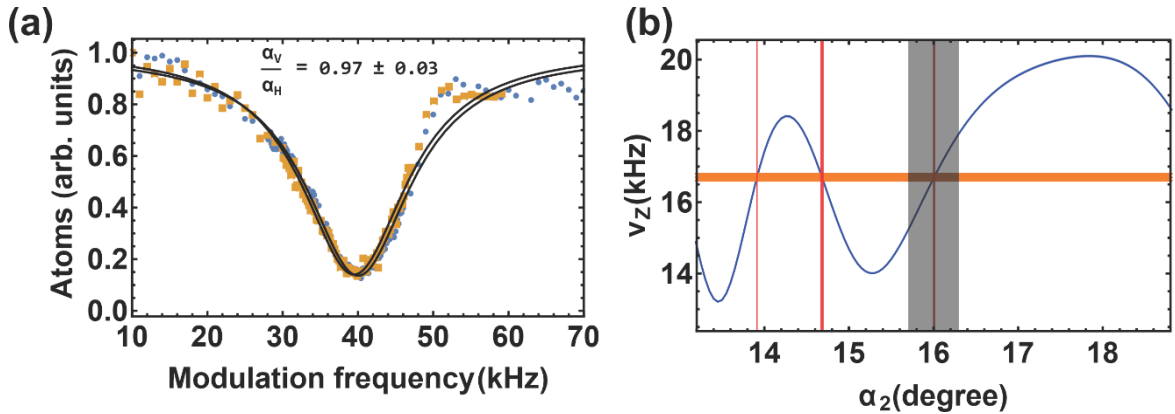

Figure 4. a) Dependence of atom loss versus the frequency of modulation in two different cases of the magnetic field. Blue dots correspond to the Oz direction of the magnetic field while orange dots correspond to the horizontal (Oy) direction of the magnetic field. Black lines represent the fit of the data with the Lorentzian function. b) Frequency $\nu_z$ from model eq.(11) versus the angle of the second beam $\alpha_2$ for the power 4.0 W per beam. The first beam has an angle of 16°. The blue solid line is the result of the calculation for trap frequency. The underlying shaded area represents the trust region coming from experimental uncertainties [38]. The orange rectangle shows the frequency of 16,699±94 Hz measured experimentally. The black rectangle represents the area, which is excluded from the model due to the experiment with BEC diffraction. The red transparent rectangles show the Figure 4b shows the experimentally found trap frequencies along with the simulated frequencies as a function of an angle $\alpha_2$ (see Figure 1a) between the $k_2$ beam of the OL and horizontal plane. The frequency $\nu_z$ was calculated for the power of 4.0 W per beam using scalar and tensor polarizabilities from [35] and the model (6)-(11) [38]. Oscillations of the calculated trap frequency are caused by the phase between the fields $\vec{E}_{10}, \vec{E}_{20}$ and $\vec{E}_{11}, \vec{E}_{21}$, respectively, as it depends on the angle of incidence on the window $\alpha_2 - 16°$. The oscillatory behavior of the model leads to 3 possible solutions for the angle $\alpha_2$, only one of which is excluded by the BEC experiment.

The polarizability of thulium was measured with finite precision. Mostly, the systematic uncertainty of the experiments leads to the uncertainty of the calculated value of polarizability. Nevertheless, since polarizability was measured in the same setup and was measured via measurements of trap frequencies, thus, in reality, the comparison of trap frequencies with trap frequencies at the same setups and therefore the systematic uncertainty of the polarizability measurements may be excluded from consideration. In this logic, the allowed angles $\alpha_2$ are $\alpha_2 = 13.83 \pm 0.15$, $\alpha_2 = 14.75 \pm 0.13$.

If systematic uncertainty must be taken into account, it will significantly reduce the precision with which the angle could be found, but it could be corrected by the measurement of trap frequency with a known beam at the same setup.

## V. CONCLUSIONS

BEC was successfully used to diagnose the parameters of an OL using the Kapitsa-Dirac effect. It was determined that one of the lattice beams had self-reflection, interfering with the main lattice. The period of this lattice, as well as the period of the main lattice, were extracted from BEC diffraction images and were congruent with expectations. The correctness of lattice picture reconstruction was checked via polarizability measurements and was found to be congruent with the previously measured value. Moreover, the angle of incidence of one of the beams was restricted to two discrete values using the combination of BEC diffraction data and trap frequency measurements.

## VI. ACKNOWLEDGMENTS

This work was supported by Rosatom in the framework of the Roadmap for Quantum computing (Contract No. 868-1.3-15/15-2021 dated October 5, 2021).